# Generation of superhelical time-varying OAM beam with space-time-coding digital metasurface

Jingxin Zhang[1], Peixing Li[2], Alex M H Wong[2]*, Jensen Li[1]*

[1] Department of Physics, The Hong Kong University of Science and Technology, Clear Water Bay, Kowloon, Hong Kong, China
[2] State Key Laboratory of Terahertz and Millimeter Waves, Department of Electrical Engineering, CityUniversity of Hong Kong, Hong Kong
* alex.mh.wong@cityu.edu.hk, jensenli@ust.hk

## Abstract

The recently proposed extreme-ultraviolet (EUV) beams with time-varying orbital angular momentum (OAM) realized by high-harmonic generation (HHG) provides extraordinary tools for quantum excitation control and particle manipulation. However, such an approach is not easily scalable to other frequency regimes. Here, we design a space-time-coding digital metasurface operating in microwave frequencies to experimentally generate different modes of time-varying OAM. We also introduce a concept of superhelicity of time-varying OAM with a higher-order twist of the wavefront structure, which can be further realized by reflection phase profiles nonlinear in time through the metasurface. By developing a two-probe measurement technique, we dynamically map the OAM field pattern varying in time. The proposed superhelical time-varying OAM beam paves a way for particle manipulations and OAM multiplexing communications.

## Introduction

Electromagnetic (EM) waves carry both linear momentum and angular momentum. Particularly, the latter can be decomposed into spin angular momentum (SAM) and orbital angular momentum (OAM)[1]. The SAM is related to the polarization state of EM waves, with two values of $+1$ and $-1$ for two opposite directions of circular polarization. OAM is associated with helical wavefront of EM waves and a transverse phase profile of $\exp(il\theta)$, where $l$ is an unbounded integer defining the topological charge (OAM mode) and $\theta$ is the azimuthal angle[2]. OAM beams with distinct $l$ are mutually orthogonal, allowing them to carry information and to be multiplexed and transmitted along the same beam axis[3-7]. Hence, the channel capacity and spectral efficiency can be potentially increased tremendously. OAM multiplexing has enormous application potential in the field of communications, even if many practical technical difficulties still need to be considered. Besides, the OAM beams have a topological phase singularity along the beam axis, which leads to a vanished intensity at the axial center field. Such properties of OAM beams allow further applications in particle manipulation such as optical tweezers, lattices, and centrifuges [8-10].

In general cases, OAM-carrying beams keep constant OAM mode $l$ with phase profile of $\exp(il\theta)$ along the propagation direction. The radius of the dark central spot caused by topological phase singularity along the OAM beam axis remains unchanged. Recently, several studies have been conducted to generate a spatiotemporal optical vortex (STOV) whose topological charge has spatial dependence in the propagation process[11-16]. Such kind of beams manifests different topological charges at different positions, yet the topological charge remains unchanged with time at the same location. In 2019, Rego proposed a kind of extreme-ultraviolet (EUV) beams with time-varying OAM arising in high-harmonic generation (HHG) driven by time-delayed pulses with different OAM[17]. The topological charge is varying continuously between two values of $l1$ and $l2$ (i.e., $l1 < l(t) < l2$) with respect to time, which means the time-varying OAM beams have a time-varying phase profile of $\exp[il(t)\theta]$. This property was termed as self-torque of light. Such kind of dynamic-OAM beams finds unprecedented applications in imaging magnetic, launching selective and chiral excitation of quantum matter[18], imprinting OAM centrifuges[19], or manipulating the OAM dichroism of nanostructures[20]. However, such an HHG process is not easily applied in other frequency regimes. More generic approaches are required for the time-varying OAM beam generation.

Metasurfaces, artificially engineered structures consisting of sub-wavelength unit cells, have the unprecedented capability of manipulating the EM waves[21-23], which may provide an alternative approach for the time-varying OAM beam generation. Sedeh carried out a theoretical analysis for time-varying OAM generation by designing an optical metasurface with azimuthal modulation frequency gradient[24]. However, the experimental implementation of such a scheme is still elusive for time-modulating a metasurface operating in the optical regime. On the other hand, the recently proposed space-time-coding digital metasurface[25-28] in microwave regime uses discrete bias voltages to lower the demand of high modulation frequency, which may enable the design of a space-time-coding digital metasurface for the time-varying OAM beam generation. Moreover, for experimental verification of the time-varying OAM beam generation, we also need to observe the time-varying OAM field pattern at different time instances. Both the amplitude and phase field patten of each individual pixel need to be measured simultaneously. While phase information is easier to obtain for microwave and Terahertz experiments using a scanning probe, the modulation cycle will not be aligned for different pixels when the probe moves. All these factors make experimental evidence for time-varying OAM beam generation using such approaches still elusive.

In this paper, we experimentally construct and observe time-varying OAM beams. In addition, a digital approach for time-modulation allows extra flexibility for us to further explore more degrees of freedom in constructing time-varying OAM beams. We explore the concept of superhelical time-varying OAM beam by a programmable space-time-coding digital metasurface. Superhelicity of time-varying OAM is here defined as a higher-order twist of the wavefront structure, which can be realized by creating nonlinear time-varying reflection phase through the metasurface. As shown in Fig. 1a, we design a space-time-coding reflective metasurface working at 11 GHz, in which each unit cell is controlled by digital coding signal from field-programmable gate array (FPGA) system. The wavefront of the reflected wave can be engineered to generate a superhelical time-varying OAM beam with topological charge $l(t)$ changing from $l = -4$ to $l = 4$ periodically in time. Besides, the wavefront profile shows a higher-order twist structure compared with the OAM beam with constant $l$. By utilizing a two-probe measurement technique, we can map the time-varying OAM field pattern dynamically to demonstrate our theoretical expectations. The superhelicity of the time-varying OAM beam can be regarded as an additional degree of freedom in OAM wavefront structure design, which can be useful in particle manipulations and OAM multiplexing communications.

## Results

**Theory underlying the superhelical time-varying OAM generation**

We consider a space-time-coding digital metasurface divided by N azimuthal sections. The reflection coefficient of the metasurface depends on azimuthal angle $\theta$ and time $t$, which can be expressed as

$$r(\theta, t) = A \exp\left[iN\frac{t}{T}(\theta + w \cdot 2\pi\frac{t}{T})\right] \tag{1}$$

where only phase modulation is considered and the amplitude A can be seen as a constant. By defining $l(t) = Nt/T$ and $\delta\theta(t) = 2\pi(t/T)$, Eq. (1) can be rewritten as

$$r(\theta, t) = A \exp\left[il(t)(\theta + w \cdot \delta\theta(t))\right] \tag{2}$$

Under the excitation of a monochromatic wave $E_i(t) = E_0 \exp(i\omega_0 t)$, where $\omega_0$ is the carrier frequency, the reflected wave can be expressed as

$$E_r(\theta, t) = E_0 A \exp[i\omega_0 t] \exp\left[il(t)(\theta + w \cdot \delta\theta(t))\right] \tag{3}$$

As it can be seen, we obtain an OAM beam with time-varying topological charge $l(t)$. The time-varying term $w \cdot \delta\theta(t)$ leads to the resultant phase change nonlinear in time t, which can be seen as higher-order twist of the wavefront structure of the time-varying OAM beam as shown in Fig. 1a. Different color mean different phase values. When we join up the zero phase (cyan color) at different time instances, the wavefront structure shows up a twist for a non-zero $w$ (value -1 in this example). We call such a property as superhelicity of a time-varying OAM beam, described by the winding number $w$. Zero $w$ returns to the previously defined time-varying OAM[24]. Considering $l(t) = Nt/T$, at discrete time $t = 0, T/N, 2T/N, ....$, the reflected wave has topological charge $l = 0, 1, 2, ....$, increasing with time t. Practically, the metasurface can only be divided into finite N pieces to get discrete $\theta = m\Delta\theta$ with $\Delta\theta = 2\pi/N$, where $m = 1, 2, ..., N$. It is easy to find that topological charge $l$ is equivalent with $l + N$ due to

$$\exp(i(l + N)m\Delta\theta) = \exp(i(lm\Delta\theta + 2\pi m)) = \exp(ilm\Delta\theta) \quad (4)$$

revealing that the topological charge $l$ will be repeating with a period of $T$.

To generate such a superhelical time-varying OAM beam by space-time-coding digital metasurface, we adopt discrete phase change realized by digital coding sequence instead of using continuous phase modulation. As shown in Fig. 1b, we divide the metasurface into N=8 pieces and 3-bit coding digits "1", "2", "3", "4", "5", "6", "7" and "8" are used to represent 8 phase states with an interval of 45°. The space-time-coding scheme is designed according to Eq. (1) but with discrete time t. The black stars in Fig. 1b mark the azimuthal positions of coding digit "1" (0° phase) showing the superhelicity of the time-varying OAM with $w = -1$.

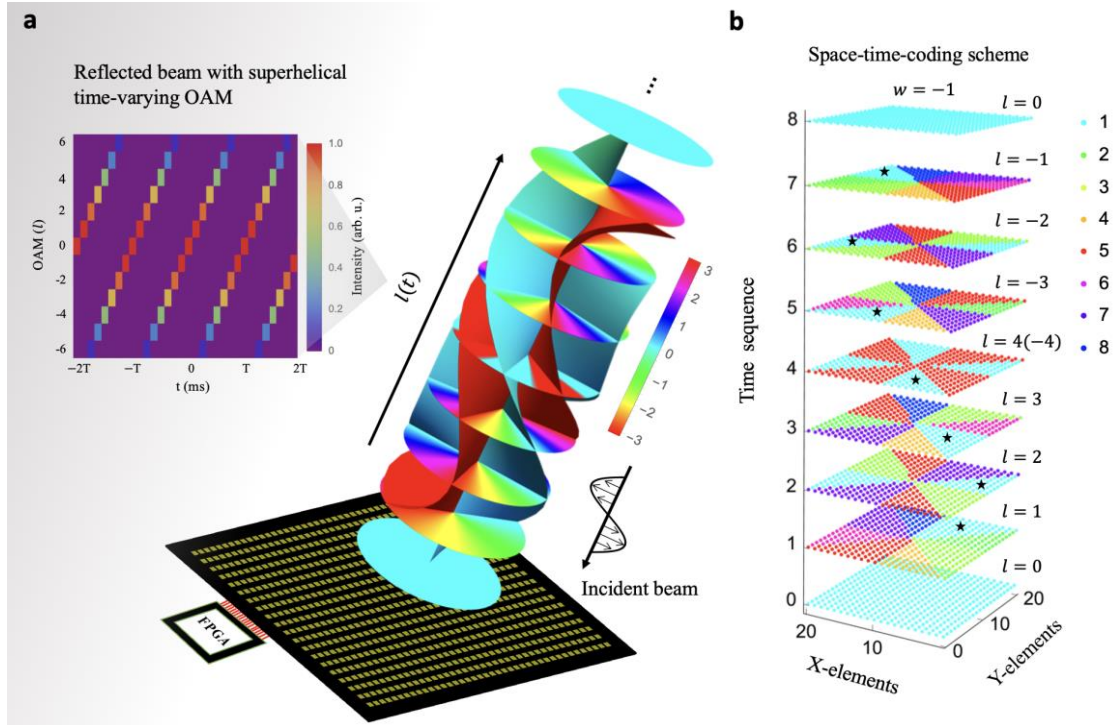

**Fig. 1 Superhelical time-varying OAM beam generation by a space-time-coding digital metasurface**. **a** Conceptual illustration of superhelical time-varying OAM beam generation by a space-time-coding reflective metasurface, where $l(t)$ illustrates the temporal variation of the topological charge in the generated OAM beam. The blue and red curved surfaces show the wavefront profile of 0 and π phase, respectively. **b** The digital coding scheme of the proposed space-time-coding digital metasurface, in which the 3-bit coding digits "1", "2", "3", "4", "5", "6", "7" and "8" represent the 0°, 45°, 90°, 135°, 180°, 225°, 270° and 315° phase responses of the reflected wave.

## Space-time-coding digital metasurface

To experimentally demonstrate the above theories, we design and fabricate a space-time-coding digital metasurface consisting of $20 \times 20 = 400$ unit cells for the superhelical time-varying OAM beam generation as shown in Fig. 2a. The size of the unit cell is 10mm×10mm, which is close to one-third of the wavelength at the working frequency of 11GHz. Varactor diodes are embedded in each unit cell of the metasurface as active components to obtain time-varying reflection phase electronically. The pins on the side of the metasurface are used for the connection with FPGA system to realize bias voltage control. These pins are rearranged into 8 groups by the biasing network on the bottom layer (see Supplementary Fig. S1) to control the 8 azimuthal sections for the digital coding scheme as shown in Fig. 1b.

The 3D structure of each unit cell of the proposed metasurface is illustrated in Fig. 2b. The geometrical parameters of the unit cell are chosen as: $h_1$ = 0.813 mm, $h_2$ = 0.1 mm, $h_3$ = 0.305 mm, $\varnothing_1$ = 1.2 mm, $\varnothing_2$ = 0.8 mm, $s$ = 0.6 mm, $a$ = 10 mm, $w$ = 5 mm, $l$ = 4 mm, $g$ = 1 mm, $p$ = 1.4 mm, $d$ = 0.3 mm. The unit cell is composed of three copper layers printed on two substrate layers (Rogers 4003C, $\varepsilon_r = 3.55$, $tan\delta = 0.0027$) and a bonding layer (Rogers 4450F, $\varepsilon_r = 3.52$, $tan\delta = 0.004$). On the top layer, two rectangular metal patches are connected by a varactor diode (MAVR-000120-14110P). One patch is connected to the middle layer through metallic via−. Another patch is connected to the bottom layer through via+, which is electrically isolated from the middle layer by a hollowed ring. The middle layer connects to negative "−" electrode and the bottom layer connects to positive "＋" electrode, providing the bias voltage to the varactor diode through the two metallic vias.

The capacitance of the varactor diode will change with the bias voltage, leading to the frequency shift of dipole resonance in the metasurface, so that the reflection phase can be varied dynamically for phase modulation. In practical implementation, we prefer to apply discrete phase control instead of continuous phase control to lower the demand for high modulation speed supported by FPGA system. We use 8 discrete phase states from 0° to 360° with an interval of 45° to realize 3-bit digital coding in our paper. To realize 3-bit reflection phase control at the frequency of 11GHz, 8 bias voltages are selected from U1 to U8, and the simulated reflection amplitude and phase at each bias voltage are shown in Fig. 2c, d. At the operating frequency of 11GHz, we acquire 8 different reflection phases with a 45° gradient covering 315° range, which can be used as 3-bit coding digits "1", "2", "3", "4", "5", "6", "7" and "8".

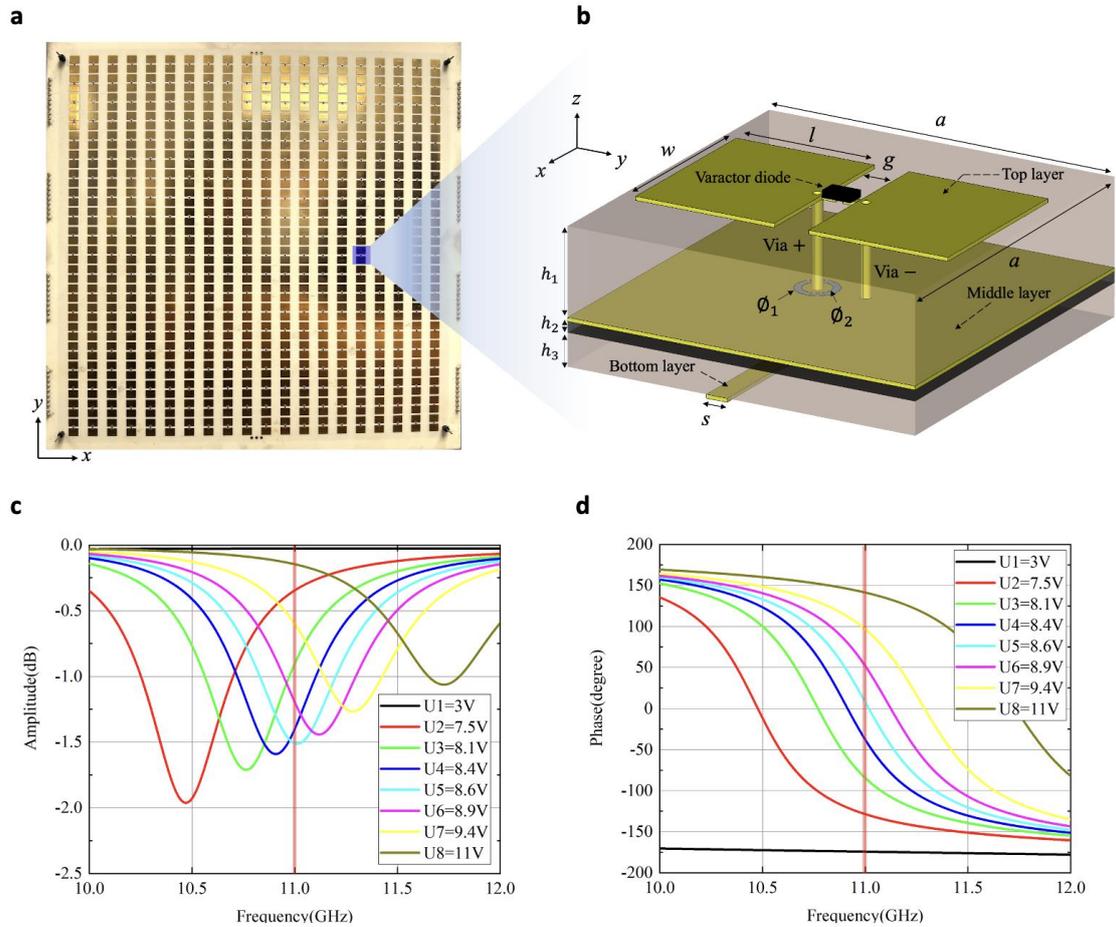

**Fig. 2 Metasurface design and reflection response.** **a** The photo of the fabricated

space-time-coding digital metasurface. **b** The detailed geometrical parameters of the unit cell structure **c, d** The simulated reflection amplitude and phase of the metasurface at different bias voltages, where the highlighted area represents the operating frequency at 11GHz.

**Dynamic field pattern measurement of the superhelical time-varying OAM beam**

For experimental demonstration of the superhelical time-varying OAM beam generated by the space-time-coding digital metasurface, we directly measure the time-varying field pattern including amplitude and phase information at various time instances. To map the fields in the time-domain, a network analyzer can be used in the time-sweep mode with time-window much shorter than the modulation cycle. However, the measured S-parameters can be at any instance within one modulation cycle, the fields probed at different positions cannot be compared with each other. Here, we add one more probe R as reference (in addition to the signal probe S) whose position is fixed. Then, the result from the reference probe R does not vary against different signal probe positions except a time shift, which can now be used to align the S-parameters between signals probed at different positions (see the schematic in Supplementary Fig. S2 and S3 for more details).

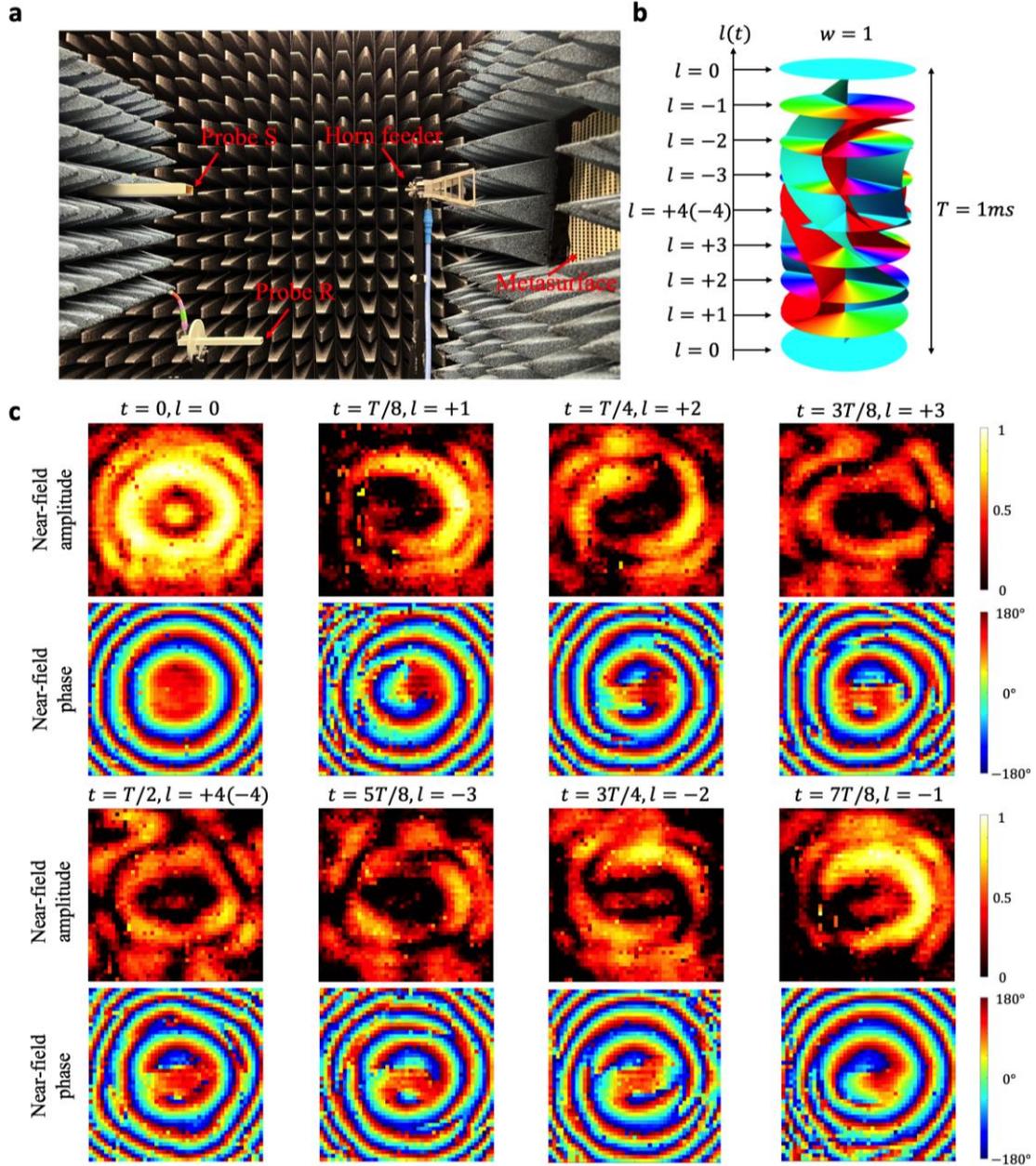

**Fig. 3 Dynamic measurement of superhelical time-varying OAM field pattern by using two-probe technique. a** The experimental scenario of dynamic electric field pattern measurement of the generated superhelical time-varying OAM beam. **b** The schematic of the expected superhelical time-varying OAM changing from $l = -4$ to $l = 4$ with winding number $w = -1$. **c** The measured near-field amplitude and phase patterns of the superhelical time-varying OAM beam at different instants of time in one period of $T = 1ms$, and the near-field pattern range is $0.6\text{ m} \times 0.6\text{ m}$.

In our experiment, as shown in Fig. 3a, a prototype of metasurface consisting of

$20 \times 20 = 400$ elements is fabricated and 400 varactor diodes (MAVR-000120-14110P) are soldered on the metasurface by the surface mounting technology (SMT). Each azimuthal section is applied with 3-bit digital coding sequence as shown in Fig. 1b, generating the superhelical time-varying OAM beam whose topological charge changes from $l = -4$ to $l = 4$ and winding number $w = -1$ with a period of 1 ms as shown in Fig. 3b. The horn feeder is placed 0.2 m away from the reflective metasurface to give a monochromatic excitation signal at 11 GHz in y polarization. Two open-end waveguide probes (OEWG, WR-90) are placed 0.44 m away from the metasurface. One probe S is moveable and controlled by a moving stage for planar near-field scanning with a scanning range of 0.6 m×0.6 m, while the other probe R was fixed at a location which is close to but outside the scanning range to receive a reference signal for synchronization processing. The horn feeder, probe R, and probe S are connected to ports 1, 2, and 3 of a network analyzer (Agilent N5230A), respectively. The stimulus condition of the network analyzer is set as continuous wave (CW) time sweep at 11 GHz, with a sweep time of 5 ms. In CW working mode, the network analyzer can measure a time-domain S-parameter waveform (complex value) with a temporal resolution of 12.5 μs in our case. By simultaneously measuring the probe R and probe S in the scanning range of 0.6 m×0.6 m, $45 \times 45 = 2025$ groups of data are obtained. By doing data post-processing for probe S at all scanning positions (see details in Supplementary Section 2), we can map the dynamic amplitude and phase field pattern of the generated superhelical time-varying OAM beam at any moment during a modulation cycle.

The Fig. 3c illustrate the measured near-field amplitude and phase patterns of the superhelical time-varying OAM beam at different instants of time in one period of T, showing the changing topological charge $l = 0, \pm1, \pm2, \pm3, \pm4$ with winding number $w = -1$. The field patterns consisted of $45 \times 45 = 2025$ pixels with a range of 0.6 m × 0.6 m. The change of OAM is clearly observable as the spiral wavefronts in time. To further evaluate the OAM mode purity and illustrate the superhelicity of the generated time-varying OAM beam, we calculated the OAM spectrum including the intensity and phase by using the following equation

$$A_l(t) = \frac{1}{2\pi} \int_0^{2\pi} \psi(\theta, t, r_0) \exp(-il\theta) \, d\theta, \tag{5}$$

where $\psi(\theta, t, r_0)$ denotes the measured time-varying OAM field in Fig. 3c, $r_0$ is center radius. We note that due to the discretization error of 8 pieces in the metasurface and the blocking effect of the horn feeder, we do not use the field near the center and $r_0$ has to be sufficiently large. Then the OAM intensity and OAM phase are defined as $I_l(t) = |A_l(t)|^2$ and $\varphi_l(t) = \arg(A_l(t))$, where $\arg(z)$ is the argument of a complex number z. The calculated OAM spectrums at different instants of time are shown in Fig. 4. As it can be seen, the OAM intensity at desired topological charges $l$ are relatively dominant compared with the parasitic ones, proving it has OAM being linear in time. The OAM phase $\varphi_l(t)$ carries the orientation angle information of the time-varying OAM beam at the $l$ mode. We only focus on the $\varphi_l(t)$ with dominant intensity $I_l(t)$ as marked with white stars in the Fig. 4. The OAM phases $\varphi_1(T/8)$, $\varphi_2(T/4)$, $\varphi_3(3T/8)$, $\varphi_4(T/2)$ $\varphi_{-3}(5T/8)$, $\varphi_{-2}(3T/4)$, $\varphi_{-1}(7T/8)$ are obtained as 235.5°, 51.6°, 309.4°, 298.1° 252.3°, 128.1° and 78.1°, respectively. We define the azimuthal angle for the piece with coding digit "1" (0° phase) in Fig. 1b as $\theta_s(t)$ that can be expressed as $w \cdot 2\pi k(t)/N$, where $k(t) = Nt/T$, and $w$ is the winding number of the time-varying OAM beam. The OAM phase $\varphi_l(t)$ can be fitted by

$$\varphi_l(t) = l(t)(\gamma \theta_s(t) + \alpha) + \beta \tag{6}$$

where the $t_0$ is initial moment, α and β are fitting parameters. The predicted OAM phases $\varphi'_1(T/8)$, $\varphi'_2(T/4)$, $\varphi'_3(3T/8)$, $\varphi'_4(T/2)$ $\varphi'_{-3}(5T/8)$, $\varphi'_{-2}(3T/4)$, $\varphi'_{-1}(7T/8)$ are 234°, 50°, 298°, 258°, 250°, 138°, 98° with $\alpha = -40°$ and $\beta = 130°$, and we note that there is a fitting parameter $\gamma \cong 0.8$, which associates to the inaccurate translation from voltage to the reflection phase, the finite number of sectors, and also the dynamic range of the voltage cannot cover fully $2\pi$. Nevertheless, the results prove the field pattern has a superhelicity of $w = -1$.

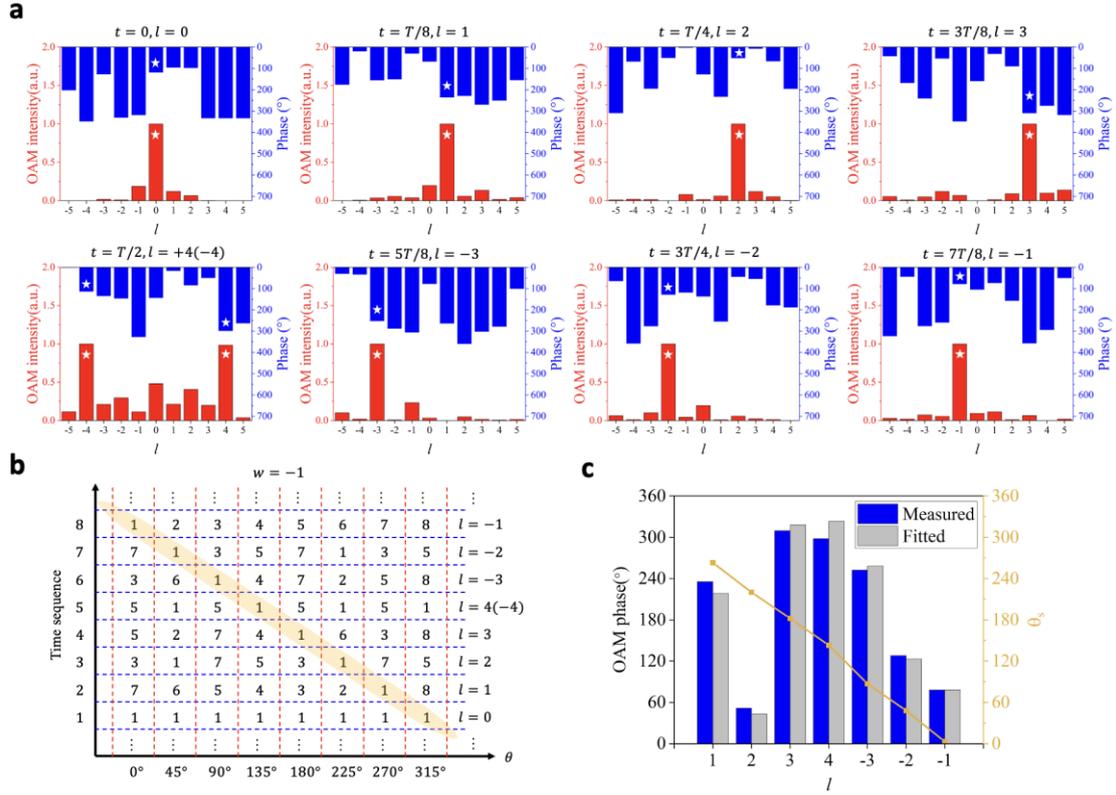

**Fig. 4 Spectrum analysis of superhelical time-varying OAM. a** The OAM intensity and phase spectrum of the measured superhelical time-varying OAM beam at different instants of time. **b** The coding book showing the azimuthal positions of each phase state. **c** The comparison of measured and fitted OAM phases for different OAM mode.

## Discussion

For experimental validation of the time-varying OAM beam generation, one stumbling block is the high-speed modulation requirement for arbitrary continuous modulation waveform, the other is the dynamic measurement of the time-varying amplitude and phase field pattern. In this paper, we design a space-time-coding digital metasurface working in microwave frequency regime. Instead of utilizing continuous modulation signal, we apply a discrete digital coding signal supplied by FPGA system to reduce the requirement for modulation speed. And the coding sequence can be arbitrarily designed owing to the flexibility in digital signal processing. Besides, we have visually observed the time-varying field pattern including amplitude and phase information to verify the OAM mode is time-varying according to our theoretical prediction. However, the traditional near-field scanning system in microwave experiment (conventionally with one probe) cannot fulfill our demands because each measurement of the probe during the scanning process is not aligned with the modulation cycle. Then a two-probe measurement technique is proposed to map the time-varying OAM field pattern dynamically, and we can successfully observe the OAM amplitude and phase pattern varying in the time domain, which enables the experimental validation of time-varying OAM generation and other time-varying scenarios. Moreover, due to the flexibility of the space-time-coding digital metasurfaces, such approach allows us to explore higher order wavefront structure of time-varying OAM beams. We have introduced the winding number $w$ for the time-varying OAM beams. It leads to the concept of superhelicity of time-varying OAM with a higher-order twist of the wavefront structure.

In summary, we experimentally generate a superhelical time-varying OAM beam with topological charge changing from $l = -4$ to $l = 4$ periodically by a space-time-coding digital metasurface in microwave regime. The concept of superhelical time-varying OAM beam with a higher-order twist of the wavefront structure is proposed by designing reflection phase profiles nonlinear in time through the metasurface and has been realized. The dynamic image of OAM field pattern being varied in time domain are mapped by the two-probe measurement technique. The proposed superhelical time-varying OAM beam generation may find its application potentials in particle manipulations and OAM multiplexing communications.

**Acknowledgment**

This work is supported by the Hong Kong Research Grants Council with project numbers R6015-18 and C6012-20G.